\documentclass{PoS}

\usepackage{graphicx}
\usepackage{xspace}

\newcommand{\url}[1]{\href{#1}{\tt #1}}
\usepackage{amsmath}
\usepackage{amssymb}

\newcommand{\as}{\alpha_{\text{\sc s}}}

\newcommand{\order}[1]{{\cal O}\left(#1\right)}
\newcommand{\cpp}{\texttt{C++}\xspace}

\title{Perturbative QCD for the LHC}

\ShortTitle{Perturbative QCD for the LHC}

\author{Gavin P. Salam
  \\
  CERN, Department of Physics, Theory Unit, CH-1211 Geneva 23, Switzerland\\
  Department of Physics, Princeton University, Princeton, NJ 08544, USA\\
  LPTHE, UPMC Univ.~Paris 6 and CNRS UMR 7589, Paris, France\\
  E-mail: \email{salam@lpthe.jussieu.fr}}

\abstract{%
  These proceedings discuss some of the highlights of recent research
  in perturbative QCD as it relates to the LHC.
  Topics covered include the new generation of Monte Carlo event
  generators, the revolution that is occurring in NLO calculations,
  progress towards NNLO predictions and developments in the definition
  and use of jets.\medskip

  This writeup is dedicated to the memory of Thomas Binoth and Ulrich
  Baur.  }

\FullConference{35th International Conference of High Energy Physics - ICHEP2010,\\
		July 22-28, 2010\\
		Paris France}

\begin{document}

\section{Introduction}

As the LHC programme gets underway, it is timely to examine the status
of our QCD tools.
One may ask whether they have reached the degree of sophistication
that we expected of them at this stage.
It is quite an achievement for the community to be able to say that
the answer to this question is ``yes''.
A second question is whether our QCD tools have achieved the degree of
sophistication that is necessary for fully exploiting the LHC over the
years to come.
Here the answer will be more nuanced: there is certainly still ample
scope for progress.

To provide a context for our discussion of QCD, it is perhaps worth
briefly recalling some of the roles played by QCD at a hadron collider
whose key design aims are not to study QCD, but rather to discover the
Higgs boson and search for physics beyond the standard model.

There are essentially two ways of making discoveries at the LHC.
On one hand, an experiment may measure some kinematic distribution and
see a discrepancy relative to the standard-model expectation.
It can only be labelled a discovery if one has sufficient confidence
in the standard model prediction, inevitably involving many aspects of
QCD, such as parton distribution functions, matrix elements, parton
showers, etc.\footnote{This is true even with many ``data-driven''
  methods for estimating backgrounds, since they often rely on
  QCD-based extrapolations from the region where one has a good
  measurement of the background to that where one suspects the
  presence of the signal.}
Alternatively, discovery may come through the observation of a
distinct kinematic structure, such as an invariant mass peak (or edge,
in the presence of unmeasured particles).
At first sight, QCD might have less of a role to play here; however,
an understanding how QCD works can make it possible to reduce the
backgrounds, and sharpen the kinematic structure of signal, allowing
it to emerge more convincingly.
Furthermore, as and when discoveries are made, QCD will also be crucial in
extracting information about the new objects that have been found:
their couplings, masses, spins, etc.

In these proceedings we will examine several areas of perturbative QCD
that have seen major milestones in the past year or two. The first
such area is that of Monte Carlo event generators.

\section{Parton-shower event generators (Monte Carlos)}

It is almost inconceivable to think of the LHC experiments working
without Monte Carlo (MC) programs such as
Pythia~\cite{Sjostrand:2006za}, Herwig~\cite{Corcella:2000bw} and
Sherpa~\cite{Gleisberg:2003xi}, which output detailed simulated pp
collision events.
The immense preparatory effort for the LHC would not have been
possible without these tools, be it for the investigation of physics
potentialities, or the simulations of detector response;
nearly all of the data shown by LHC experiments at ICHEP 2010 (and
since) have been accompanied by comparisons to MC simulations, most
often in amazingly good agreement;
and as the experiments move towards producing results at ``particle
(hadron) level'', i.e.\ corrected for detector effects, MC simulations
will always be central in determining those corrections.

The core of the code base for two of the main MC tools, Pythia (v6)
and Herwig, dates to the 1980's (early versions of Sherpa also used
portions of Pythia code) and is written in Fortran~77, a language that
strains to adapt to the sophistication that these programs have
reached today.
This prompted an effort across the Monte Carlo community, initiated
almost a decade ago, to rewrite the programs in \cpp.
Aside from the magnitude of the task of rewriting $60-80,000$ lines of
code, many questions of \cpp design needed to be thought through
carefully, to ensure that the new code remained maintainable over the
lifetime of the LHC.
One of the milestones of the past couple of years is that, coinciding
with the start of the LHC, the new versions of these programs,
Pythia~8.1~\cite{Sjostrand:2007gs}, Herwig++~2.4~\cite{Bahr:2008pv}
and Sherpa~1.2~\cite{Gleisberg:2008ta}, are now available and mature
enough for production use, including all core features needed for
complete hadron-collider analyses, such as simulation of the multiple
interactions.

The work towards the \cpp generators has not simply been a question of
rewriting old code in \cpp (for a comprehensive review, see
ref.~\cite{Buckley:2011ms}).
For example, Pythia has acquired a new $p_t$ ordered shower as its
default~\cite{Sjostrand:2004ef} (the old Fortran virtuality-ordered
shower is no longer
available in the \cpp version); it also includes numerous developments
related to multiple interactions, e.g.~\cite{Corke:2010yf} and its
modularity has already been exploited to allow the inclusion of an
alternative shower~\cite{Giele:2011cb}.
Herwig has updated its angular-ordered shower~\cite{Gieseke:2003rz},
including better treatment of massive particles, and it now
incorporates a native multiple interaction model~\cite{Bahr:2009ek}.
Sherpa did not have a corresponding Fortran version, however it has
seen a number of significant developments in the past couple of years,
most notably the switch to a dipole shower, and efficient multi-leg
matrix elements (COMIX~\cite{Gleisberg:2008fv}, used together with
CKKW \cite{Catani:2001cc} matching to the parton shower).
Other progress in the generators includes more extended BSM support
and inclusion of NLO corrections for a broad variety of processes (as
discussed below).

Overall, it is time for this new generation of codes to undergo
extensive stress testing by the experiments, the last major step on
the way to their becoming the Monte Carlo workhorses for the duration
of the LHC.

\section{The NLO revolution}

While Monte Carlo event generators give fine-grained predictions
about QCD final states, it is not always simple to systematically
improve their accuracy.
The most straightforward systematically improvable calculational
approach of QCD at high energies is to use a perturbative
approximation, involving a series expansion in the strong coupling
$\as$, i.e.\ cross sections are written $\sigma = c_0 + c_1 \as + c_2
\as^2 + \ldots$, so that an improvement in accuracy is obtained
``just'' by calculating one further coefficient in the series.

At the momentum scales of relevance for LHC, $\as \simeq 0.1$ and one
would expect a leading order (LO) calculation, one that includes just
the first non-zero term of the series, to be accurate to within about
$10\%$.
Yet widespread experience shows that this is seldom the case, with
next-to-leading order (NLO) corrections often modifying cross sections
by a NLO/LO ``$K$ factor'' ratio of two (for example for Higgs
production~\cite{Dawson:1990zj,Djouadi:1991tka,Harlander:2002wh} or
$Wb\bar b$~\cite{Ellis:1998fv,FebresCordero:2006sj}).
In some situations, in which a new channel opens up at NLO,
$K$-factors can be much larger, even
$\order{100}$~\cite{Bauer:2009km,Denner:2009gj,Rubin:2010xp}.
These NLO enhancements are potentially important because, for example,
in searches for supersymmetry the ``signal'' of supersymmetry is often
just a factor $\order{5}$ excess of the data over the expected
background~(e.g.~\cite{ATLAS:2011hh}), the latter nearly always being
calculated at LO.
%
How, then, do we determine whether an excess of data compared to LO is
an actual signal or simply a background with an unexpectedly large,
$K$-factor that has yet to be calculated?

Part of the answer is that experiments attempt to constrain the
$K$-factor in regions of phase-space expected to be signal-depleted. 
However extrapolations to possible signal regions still usually
involve LO tools,\footnote{An interesting distinction here is between
  simple LO, and matched matrix-element plus Monte Carlo samples
  involving multiplicities that go beyond the strictly LO
  process~\cite{deAquino:2011ix}, which can account for the appearance
  of new higher-order channels and reproduce some NLO $K$-factors.}
and the known cases with the largest $K$-factors usually also lead to
strong kinematic dependence of the NLO correction.
In such situations, therefore, it would be reassuring to have an
actual NLO calculation.
The difficulty is that many new physics signals involve quite complex
backgrounds.
For example, in pair production of gluinos, each gluino might decay to
a anti-quark and squark, with each squark decaying to a quark and an
(invisible) neutralino, which gives missing energy.
One of the backgrounds in this case is then four-jet production in
association with a $Z$-boson that decays to neutrinos, which is too
complex a process for there to be any NLO calculations of it yet.

One way to quantify the difficulty of a NLO calculation is in terms of
the total number of outgoing ``legs'' (partons and electroweak bosons
all count as legs).
The first NLO calculation was for a $2\to1$ process, Drell-Yan
production, in 1979~\cite{Altarelli:1979ub}.
It took almost ten years before any $2\to2$ processes got calculated,
with several results appearing in the late 1980's and early 90's
(e.g.\ heavy-quark
pairs~\cite{Nason:1987xz,Altarelli:1988qr,Beenakker:1988bq}, dijet
production~\cite{Aurenche:1986ff,Aversa:1988vb}, and vector-boson plus
jet~\cite{Arnold:1989ub,Giele:1993dj}).
Another ten years passed before a $2\to3$ process was calculated, with
$Wb\bar b$ in 1998 \cite{Ellis:1998fv} and 3-jet and $W$+2-jets
calculated a couple of years later~\cite{Nagy:2001fj,Campbell:2002tg}.

Given the motivation from the expected startup of LHC, at this point
an almost industrial effort got underway to calculate all $2\to3$
processes of interest for LHC and to open the frontier towards $2\to
4$ processes (bearing in mind that that background we mentioned above
was a $2\to5$ process), guided by a document known as the Les Houches
wishlist~\cite{Bern:2008ef}.
Roughly in line with the rule-of-thumb of a 10-year interval for
calculation an extra leg,
the first $2\to4$ calculations have appeared in the past couple of
years: $W$+3\,jets~\cite{Ellis:2009zw,Berger:2009ep}, $Z$+3-jets
\cite{Berger:2010vm}, $t\bar t b\bar b$
\cite{Bredenstein:2010rs,Bevilacqua:2009zn}, $t\bar
t$+2-jets~\cite{Bevilacqua:2010ve},
$W^\pm W^\pm$+2-jets~\cite{Melia:2010bm}, $WWb\bar
b$~\cite{Denner:2010jp,Bevilacqua:2010qb}, with progress also on $b\bar b b\bar
b$~\cite{Binoth:2009rv} (and a result for
$e^+e^-\to$5-jets~\cite{Frederix:2010ne}).

While some of these results were obtained with traditional Feynman
diagrammatic
methods~\cite{Bredenstein:2010rs,Denner:2010jp,Binoth:2009rv}, the
remaining ones have taken advantage of major developments in
``unitarity-based'' methods for calculating one-loop amplitudes (which
had been the main bottleneck for new NLO results).
Originally pioneered in the mid 1990's~\cite{Bern:1994zx}, the idea
behind these methods is to sew tree-level amplitudes together to
produce loop amplitudes, equivalent to considering loop momenta
such that specific loop propagators are on-shell.
This idea was revitalised in 2004 through the use of momenta with two
timelike components~\cite{Britto:2004nc} to broaden the set of
tree-level configurations that could be usefully
assembled.\footnote{Specifically, with two timelike components (or, in
  subsequent work, with complex Minkowski momenta), it is possible to
  have a sensible 3-particle vertex with all momenta on shell and use
  this as an ingredient in building up the loop amplitude.}
To go from this result to collider predictions has been a huge
undertaking, with many important steps along the way (most have been
reviewed in~\cite{Bern:2008ef}).
If one is to highlight a single one of them, it might be the
observation that it is possible to deduce the integrated 1-loop
diagram simply by inspection of the integrand for specific
loop-momentum configurations~\cite{Ossola:2006us}.

These developments represent a revolution in NLO calculations. Not
just because of the number of $2\to 4$ predictions that they have led
to --- a corresponding effort devoted to Feynman-diagrammatic
calculations would probably have led to a similar number of results
--- but more importantly because of the prospects that they offer for
``low-cost'' automation of NLO calculations and the extension beyond
$2\to4$ processes.
Indeed, just around the time of ICHEP, the first NLO results for a
$2\to5$ process were announced, the unitarity-based (leading colour)
calculation of $W$+4-jets~\cite{Berger:2010zx}, nearly ten years ahead
of expectations from the timeline discussed above.

One caveat to be mentioned in the context of these impressive results
is that so far most of the $2\to4$ or $2\to 5$ NLO calculations are
not yet available as public codes (with the exception
of~\cite{Melia:2011gk}).
This is perhaps a consequence of the significant complexity of the
codes, which often bring together many different tools\footnote{For
  example, on one hand the 1-loop corrections, on the other hand tools
  for handling real radiation such
  as~\cite{Gleisberg:2007md,Czakon:2009ss,Frederix:2009yq,Hasegawa:2009tx}.}
and then require enormous computing time if one is to obtain a
numerically stable result.
Nevertheless, it is only once they are public, in a form that is
relatively straightforward to use, that these calculations will be
able to deliver their full value.


\subsection{NLO and Monte Carlo event generators}

While NLO calculations have the benefit of quantifiable accuracy (at
least in regions of phase-space that don't probe disparate momentum
scales), they only ever involve a handful of partons, a far cry from
the level of detail of MC parton-shower event generators, which
predict distributions at the level of hadrons.

Two main techniques have been developed over the past decade to
combine NLO accuracy with parton shower ``detail'', the
MC@NLO~\cite{Frixione:2002ik} and POWHEG~\cite{Nason:2004rx} methods.
Generally speaking, only relatively simple processes are available: at
the time of ICHEP, not even $Z$+jet or dijet production had been
publicly implemented.
That is gradually changing thanks to progress on the systematisation
and automation of both the MC@NLO~\cite{Frixione:2010ra} and
POWHEG~\cite{Alioli:2010xd,Hoche:2010pf} methods. In the POWHEG case
this helped the implementation of $Z$+jet~\cite{Alioli:2010qp},
dijet~\cite{Alioli:2010xa} and $t\bar t$+jet~\cite{Kardos:2011qa} and
even the $2\to4$ process $W^\pm W^\pm$+2-jets~\cite{Melia:2011gk},
while in MC@NLO it has been of benefit for example in extending the
range of processes available with Herwig to work also with
Herwig++~\cite{Frixione:2010ra}.

A point to be aware of is that while NLO MC implementations of,
say, $Z$ production necessarily include a correct LO (tree-level)
$Z$+jet matrix element, they had not generally been matched with
higher-order tree-level matrix elements, e.g.\ $Z$+2-jet, etc.
In contrast, it has for some time now been standard procedure to
combine LO tree-level $Z$, $Z$+jet, $Z$+2-jet, etc. matrix elements
together (CKKW and MLM methods~\cite{Catani:2001cc,Alwall:2007fs}).
Therefore users have been forced to choose between, on one hand NLO
accuracy for simple processes but with a poor description of multi-jet
events, and on the other hand low, LO, accuracy but simultaneously for
many different multiplicities.
Ultimately one would hope to have a method that provides NLO accuracy
simultaneously for a range of different multiplicities (for example,
as implemented for $e^+e^-$ in \cite{Lavesson:2008ah}, or for
hadron-collider processes without showering in~\cite{Rubin:2010xp}).
However, in the meantime, an interesting
development~\cite{Hamilton:2010wh,Hoche:2010kg} is the merging of
POWHEG and CKKW/MLM type methods to provide NLO accuracy for the
lowest multiplicity process with LO accuracy for multijet processes.

Overall, even if it is still early days, it is clear that automation
of loop calculations, automation of methods to combine NLO and parton
showers and the development of methods to merge different
multiplicities of NLO-improved parton showers, taken together would
have the potential to radically improve the quality of MC predictions.

%

\section{NNLO}

For the foreseeable future the ultimate perturbative accuracy that one
can hope to achieve is NNLO, i.e.\ corrections up to $\order{\as^2}$
relative to the dominant process.
There are two broad reasons for being interested in NNLO
corrections. One may, for example, wish to extract precision
information about standard-model couplings (as for the Higgs boson) or
parton-distribution functions from measured cross-sections.
Alternatively one may be faced with quantities where NLO corrections
are large, and NNLO is then the first order at which one can hope to
make quantitatively reliable predictions.

NNLO hadron-collider results have been available for some time now for
Higgs and vector-boson production (state-of-the-art codes are
described
in~\cite{Anastasiou:2005qj,Grazzini:2008tf,Gavin:2010az,Catani:2009sm}),
and the current frontier is NNLO accuracy for processes with coloured
final-state particles, be they heavy (top) or light (jets).

One significant recent result is the calculation of the NNLO cross
section for Higgs production in vector-boson
fusion~\cite{Bolzoni:2010xr}, making use of the ``structure function''
approach~\cite{Han:1992hr} in which one views each proton's emission
of a vector-boson as a DIS type reaction, and then separately
considers the fusion of the two vector bosons. 
This provides a NNLO result that is inclusive over the hadronic jets,
but still exclusive with respect to the vector-boson momenta.
Numerically it indicates perturbative stability relative to the NLO
prediction, with a reduction of scale uncertainties from the $5-10\%$
range at NLO, down to $2-3\%$.

The most likely candidate for the next process to be calculated at
NNLO is $t\bar t$ production. 
Among the physics motivations, one can mention the importance of the
forward-backward asymmetry: given that it is non-zero starting only
at NLO, only from NNLO will there be some quantifiable control of the
theoretical uncertainties on its prediction.
Also of interest is the potential for an extraction of the top-quark
mass by comparing the predicted cross-section (with its relatively
strong-mass dependence) to the actual measured cross
section.\footnote{It seems this method was originally proposed during
  an extensive discussion at Moriond QCD 2008. It has since been
  analysed in detail for example in~\cite{Langenfeld:2009wd}.}

As things stood a few years ago, the ingredients that were still
missing for a NNLO calculation of $t\bar t$ production were the
following: 
the two-loop diagrams for $q\bar q\to t\bar t$ and $gg\to t\bar t$;
the squared one-loop terms for $t\bar t$ production in association
with an extra parton;
and a way of performing the phase-space integration for (tree-level)
$t\bar t$+2-parton production while keeping track of the divergences,
which need to cancel with those from the 1- and 2-loop terms.

Progress (reviewed in~\cite{Bonciani:2010ue}) started with the
calculation of the high-energy limit of the two-loop $q\bar q$ and
$gg\to t\bar t$ diagrams \cite{Czakon:2007ej}.
This was followed by a numerical evaluation of the full 2-loop $q\bar
q \to t\bar t$ amplitude \cite{Czakon:2008zk} (a corresponding
approach to $gg \to t\bar t$ seems close to completion
\cite{CzakonZurichTalk}) and by various analytical results for parts
of the two
amplitudes~\cite{Ferroglia:2009ii,Bonciani:2008az}. 
The squared one-loop terms were determined in \cite{Korner:2008bn}.
Finally, the problem of integrating the (divergent) phase-space for
production of $t\bar t$+2-partons has been solved in
\cite{Czakon:2010td}.
Thus there is hope that in the reasonably near future, first full NNLO
results for top production will become available.\footnote{%
  In the meantime there has been significant work towards estimating
  the NNLO (and yet higher-order) corrections using
  threshold-resummation
  techniques~\cite{Cacciari:2008zb,Czakon:2009zw,Ahrens:2010zv,Aliev:2010zk,Kidonakis:2010dk}.
  While it is beyond the scope of these proceedings to discuss the
  detailed differences between them, it is probably fair to say that
  they do not yet provide a consensus as to the likely impact of the
  full NNLO corrections.
}

The next frontier for NNLO calculation will probably be that of
processes with one or more light jets in both the initial and final
states, e.g. vector-boson plus jet or dijet
production.\footnote{Techniques that merge NLO calculations for
  different jet multiplicities~\cite{Rubin:2010xp} can, meanwhile,
  provide a good approximation to NNLO for those observables in such
  processes that are subject to giant $K$-factors.}
The case with final-state jets only, specifically $e^+e^- \to $3-jets,
has been solved in~\cite{GehrmannDeRidder:2007bj,Weinzierl:2008iv}.
A compilation of extractions of $\as$ based on the comparison of these
NNLO results (supplemented with resummations and non-perturbative
corrections) to event-shape data has been given in
\cite{Gehrmann:2010rj}.
Interestingly, there is a noticeable spread in the results,
highlighting the fact that at levels of precision of a few percent,
hadronic final-state observables are subject to many different effects
that can contribute at the same few-percent level as $\order{\as^2}$
corrections.
Still, NNLO corrections are a class that can be controlled, helping
provide a far more constrained discussion of the overall precision of
QCD predictions.
It is therefore highly valuable that work progresses on general NNLO
methods and their extension to processes with initial-state coloured
particles (see
\cite{Glover:2010im,Boughezal:2010mc,Anastasiou:2010pw,Bolzoni:2010bt}
and references therein).

\section{Jets}

The majority of measurements that involve hadronic energy-flow at the
LHC will make use of jets.
Jets are measured with the help of a jet algorithm, which takes the
hundreds of particles measured in an experiment and combines them into
a handful of jets.
The same procedure can be applied to theoretical parton-level
calculations, with the idea that the jets obtained from parton-level
and experiment can be directly compared.

%
A problem that had plagued hadron-collider jet measurements for nearly
20 years was that the vast majority used jet algorithms that were not
infrared and collinear (IRC) safe (despite widespread discussion of
the problem, e.g.\ \cite{Huth:1990mi,RunII-jet-physics}).
IRC safety is the property that the final hard jets should be insensitive
to the additional low-energy emissions and small-angle branchings that
occur with high probability in QCD.
Without this property, the higher-order calculations discussed above
often lead to divergent answers, compromising the huge investment that
has been made in them over the past decade.

It was therefore a welcome development to see that all of the jet
measurements presented by ATLAS and CMS at ICHEP 2010 (and the
subsequent publications, e.g.~\cite{ATLAS:2010pg,CMS:2011tk}) have
used an infrared and collinear safe jet algorithm,
anti-$k_t$~\cite{Cacciari:2008gp}
(which has 
also been used by the H1 and ZEUS
collaborations~\cite{Aaron:2009vs,Abramowicz:2010ke}).
The anti-$k_t$ algorithm repeatedly recombines the pair of particles
$i$ and $j$ that has the smallest $d_{ij} =
\min(p_{ti}^{-2},p_{tj}^{-2}) \Delta R_{ij}^2/R^2$ unless a $d_{iB} =
p_{ti}^{-2}$ is smaller, in which case $i$ is labelled a jet
($\Delta R_{ij}$ is the rapidity-azimuth separation of $i$ and $j$ and
the parameter $R$ sets the minimum interjet distance).
Closely related to the much earlier $k_t$
algorithm~\cite{KtHH,Kt-EllisSoper}, it uses a different weighting of
momentum and angle to grow jets outwards from a central core, giving
``cone-like'' jets\footnote{Jets that are nearly always circular in
  the rapidity--azimuth plane; this relates to the algorithm producing
  jets whose momentum depends linearly on the distribution of soft
  particles in the jet and its vicinity~\cite{Cacciari:2008gn}, a
  property that helps make it easier to account for detector effects.}
while remaining IRC safe.
These properties, together with earlier developments that ensure good
computational efficiency in the presence of high particle
multiplicities~\cite{Cacciari:2005hq}, help make it particularly
suitable from both the experimental and theoretical points of view.

Jet finding is not simply about comparing theory and experiment, but
also about organising the huge amount of information in an event so as
to best pull out signals of particles such as the Higgs boson and
extensions of the standard model.
One kinematic regime of particular interest turns out to be that where
particles with electroweak-scale masses are produced with transverse
momenta somewhat (or far) above the electroweak scale.
There had been a handful of early investigations of this
regime~\cite{Seymour:1993mx, Butterworth:2002tt}, and in recent years
it has become clear to what extent the hierarchy of scales present at
LHC ($\sqrt{s} \gg M_{\text{EW}}$) can usefully be exploited with
suitably targeted jet methods.
Examples include: the search for new TeV-scale particles that decay to
electroweak bosons (W, Z, H) or top-quarks, which then go on to decay
hadronically (e.g.\
\cite{Butterworth:2007ke,Thaler:2008ju,Kaplan:2008ie}; more standard
jet methods were used for example in~\cite{Baur:2008uv});
the observation that in searching for hadronic decays of the
Higgs-boson (in association e.g.\ with a
$W/Z$~\cite{Butterworth:2008iy} or a $t\bar t$
pair~\cite{Plehn:2009rk}) it may be advantageous to concentrate on the
subset of events in which the Higgs boson has $p_t \gg M_H$, or indeed
that the Higgs might be discoverable first in SUSY cascade
decays~\cite{Kribs:2010hp};
and the proposal that hadronically-decaying new particles
(e.g. neutralinos and gluinos in $R$-parity violating
supersymmetry~\cite{Butterworth:2009qa,Brooijmans:2010tn} or new
scalars that appear in buried Higgs scenarios~\cite{Chen:2010wk}) may
have sufficiently distinct jet-substructure signals to be picked out
sometimes even in purely hadronic events.\footnote{There is even a
  tantalising claim of a  hint of an excess in such a channel at the
  Tevatron~\cite{Eshel:2011vs}.}
It is beyond the scope of these proceedings to discuss in detail the
many different jet techniques that have been developed for these
purposes (among those not already cited above,
also~\cite{Almeida:2008yp,Ellis:2009su,Soper:2010xk,Cui:2010km,Kim:2010uj,Thaler:2010tr,Hook:2011cq,Barger:2011wf,Hook:2011cq,Barger:2011wf,Soper:2011cr}),
and the reader is referred instead to recent
reviews~\cite{Salam:2009jx,Abdesselam:2010pt}.

\section{Conclusions}

Several major long-term LHC-QCD related projects are now approaching
maturity.
Among them we looked at the \cpp event generators Herwig\texttt{++},
Pythia~8 and Sherpa which are all now ready for mainstream use, and
are also evolving in their physics content, be it in terms of
non-perturbative ingredients such as the underlying event or more
widespread matching with NLO calculations through automation of the
MC@NLO and POWHEG methods.\footnote{
  Though space limitations prevented us from discussing parton
  distribution functions, it is worth mentioning that the NNPDF
  project~\cite{Ball:2011mu} has likewise reached maturity in the past
  year, joining CTEQ and MSTW as a global PDF fit, involving a quite
  complementary approach to the estimation of uncertainties. The
  discussion around PDFs remains very vibrant (even controversial,
  especially in the context of Higgs exclusion
  limits~\cite{Baglio:2011wn}) and for other recent progress and
  comparisons between results, the reader is referred
  to~\cite{Guzzi:2011sv,Martin:2010db,Alekhin:2011ey,CooperSarkar:2010ik,JimenezDelgado:2010nk,Alekhin:2011sk}.}

We also looked at some breakthroughs of the past couple of years.
NLO calculations, with the first $2\to 5$ result published almost 10
years ahead of ``schedule'' (i.e.\ extrapolations of past progress)
undoubtedly belong to this category.
It is probably also fair to say that jet finding has undergone a
breakthrough, on two fronts:\footnote{Though the author is perhaps too
  close to the subject to provide an unbiased view.}
on one hand, the LHC is the first hadron collider to systematically
use infrared and collinear safe jet finding, more than 30 years after
the original proposal for jet-finding by Sterman and
Weinberg~\cite{StermanWeinberg}; on the other, it has become clear
that flexibility with jet-finding methods has the potential to help
discover Higgs-boson decay channels and new physics scenarios that had
previously been thought beyond the scope of the LHC.

One of the other areas of extensive ongoing work in QCD is the quest
for high accuracy, where we discussed the progress in NNLO
calculations (space constraints prevented a discussion of
resummations).
The most imminent development will probably be the NNLO calculation of
$t\bar t$ production, with an impact not just on predictions of the
cross section, but also, possibly, on the highly topical question of
the $t\bar t$ asymmetry.

At what point might we say that QCD is ready for the LHC? 
There has been enormous progress in the past 5 to 10 years and the
goals that were set at the turn of the century have generally been met
(with one or two good surprises along the way).
Still, in many ways, the use of QCD at colliders remains a somewhat
delicate craft, one that relies on a combination of technical skill,
physical insight and extensive experience.
This is true whether one aims for the reliable prediction of complex
backgrounds, the high-precision extraction of fundamental parameters
from data or the design of analyses that make the most of QCD to help
distinguish signal from background.
We can but look forward to breakthroughs of the coming years that will
make it more straightforward to use QCD on the path to discovery.

\section*{Acknowledgements}

I am grateful to numerous colleagues for helpful exchanges and
comments, both while preparing the talk and this writeup. Among them,
Matteo Cacciari, Aude Gehrmann de Ridder, Gudrun Heinrich, Nikolaos
Kidonakis and Giulia Zanderighi.
Financial support is acknowledged from grant ANR-09-BLAN-0060.

\end{document}